\documentclass[aps,prl,twocolumn,showpacs,amsmath,amssymb]{revtex4-1}

\usepackage{graphicx}
\usepackage{color}

\definecolor{Red}{rgb}{1.0,0.0,0.0}

\def\be#1{\begin{equation}\label{#1}}
\def\ee{\end{equation}}

\begin{document}

\title{Gauge Field Localization on the Brane Through Geometrical Coupling}

\author{G. Alencar${}^a$}
\email{geova@fisica.ufc.br}
\author{R. R. Landim${}^a$}
\email{rrlandim@gmail.com}
\author{M. O. Tahim${}^{b,}$}
\email{makarius.tahim@uece.br}
\author{R.N. Costa Filho${}^a$}
\email{rai@fisica.ufc.br}
\affiliation{${}^a$Departamento de F\'{\i}sica, Universidade Federal
do Cear\'a, 60451-970 Fortaleza, Cear\'a, Brazil}
\affiliation{${}^b$Universidade Estadual do Cear\'a, Faculdade de Educa\c c\~ao, Ci\^encias e Letras do Sert\~ao Central - 
R. Epit\'acio Pessoa, 2554, 63.900-000  Quixad\'{a}, Cear\'{a},  Brazil.}

\date{\today}
\pacs{64.60.ah, 64.60.al, 89.75.Da}

\begin{abstract}

  In this paper we consider a geometrical Yukawa coupling as a solution to the problem of gauge field localization. We show that upon dimensional reduction the vector field component of the field is localized but the scalar component ($A_5$) is not. We show this for any smooth version of the Randall-Sundrum model. The covariant version of the model with geometrical coupling simplify the generalization to smooth versions generated by topological defects. This kind of model has been considered some time ago, but there it has been introduced two free parameters in order to get a localized solution which satisfy the boundary conditions: a mass term in five dimensions and a coupling with the brane. The boundary condition fix one of them and the model is left with one free parameter $M$. First we show that by considering a Yukawa coupling with the Ricci scalar it is possible to unify these two parameter in just one fixed by the boundary condition. With this we get a consistent model with no free parameters and the mass term can be interpreted as a coupling to the cosmological constant. 
  
\end{abstract}

\maketitle

In basic physics the most simple cases are those where we regard free motion, i.e., without interactions, and the general treatment to get the correct descriptions of the system is very simple. The same idea applies if we go some steps further in the direction of field theory models. The problem is that, if we choose models without interactions, we loose richness (there are lots of examples of this richness: symmetry breaking, mass generation, particle production, topological defects, etc.). Here, we would like to discuss a kind of interaction, more specifically a geometrical coupling, in order to solve an old problem in physics of extra dimensions: the localization of gauge fields in membranes. Even in its most complex formulation - the Yang-Mills fields - gauge vector theory has in $D=4$ conformal invariance and  the basic localization procedure fall into serious problems for building a realistic model because it gives non satisfactory results (in fact we are lead to non-normalized effective actions). Solutions like to include a dilaton coupling  \cite{Kehagias:2000au} where proposed, and other suggestions that a strongly coupled gauge theory in five dimensions can generate a massless photon in the brane \cite{Dvali:1996xe}.  In this sense, it is already understood how to localize the zero modes of gravity and scalar fields \cite{Randall:1999vf,Bajc:1999mh} in a positive tension brane. The main proposal of this work is, therefore, to consider a geometrical Yukawa coupling as a solution to the problem of gauge field localization. The Yukawa coupling depends on the Ricci scalar and, as we will show, it takes an important part in simplifying previous works. It enters in a mass term and breaks the gauge invariance of the model in $D=5$, but gauge symmetry is restored in $D=4$. 

It is a known fact that the localizability of the zero mode of the
gauge field, with warp factor $A(z)=-\ln(k|z|+1)$ depends on the
finitude of the integral $\int\psi^{2}dz$, where $\psi$ satisfies a Schr\"odinger equation 
\begin{equation}\label{sch}
\frac{d^2}{dz^2}\psi-U(z)\psi=m^2\psi,
\end{equation}
with the effective potential
\begin{equation}\label{potentialA}
U=\frac{1}{4}A'^{2}+\frac{1}{2}A''=k^{2}\frac{3/4}{(k|z|+1)^{2}}-k\delta(z).
\end{equation}
The delta function acts imposing a boundary condition in $z=0$.
With this, the solution of Eq. \ref{sch} is $\psi=(k|z|+1)^{-\frac{1}{2}}$
and therefore the zero mode is not localized. Many solutions to this
problem were proposed, most of them using the idea of including new
degrees of freedom, as scalar fields or the dilaton \cite{Kehagias:2000au}. Another
solution was proposed in \cite{Ghoroku:2001zu} by the inclusion of a mass
term in the potential. However, this mass term changes only the first term of the above potential and a cubic interaction involving the delta function and the gauge
field is needed. The final action is given by
\begin{eqnarray}\label{gokuaction}
&S_{A}=-\int d^{5}X\sqrt{-g}( \frac{1}{4}g^{MN}g^{PQ}Y_{MP}Y_{NQ} \nonumber \\
&+\frac{1}{2}(M^{2}+c\delta(z))g^{MN}X_{M}X_{N})
\end{eqnarray}
where $g^{MN}$ is the metrics of the Randall-Sundrum model, $X_M$ is the gauge field and $Y_{MN}=\partial_M X_N-\partial_N X_M$. After the standard manipulations, splitting the longitudinal and the
transversal parts of the gauge field we find that we have a localized,
transversal zero mode of the gauge field if $M>0$ and $c=-2k(\sqrt{1+M^{2}/k^{2}}-1)$. Therefore the model has yet one free parameter $M$.
However, the origin of this terms is left unexplained.
In this direction recently it has been proposed in Ref. \cite{Zhao:2014iqa} a model in which this kind of terms can be generated
in a covariant way by a cubic coupling with a geometrical quantity:
\begin{equation}
S_Y=-\frac{\gamma_{1} }{2}\int d^5x\sqrt{-g} Rg^{MN}X_{M}X_{N}.
\end{equation}
In the above equation $\gamma_1$ is a parameter to be fixed by the boundary conditions, leaving the model with no free parameters. However in order to obtain the desired results they use the particular configuration of fields $\partial_{\mu}A^{\mu}=0$. After some considerations about that model we can prove that this anzats in unnecessary. First of all, note that for the RS case we have $R=16k\delta(z)-20k^{2}$.
The constant term could be a problem but the
relation between the two terms of the Ricci scalar is exactly what
we need. We see, by a simple comparison with the above action (\ref{gokuaction}), that $c=-16k\gamma_{1}$
and $M^{2}=20\gamma_{1} k^{2}$ and for $\gamma_{1}=1/16$ we get the same
results. Therefore, we can solve the problem and reduce our free parameter
to just $\gamma_{1}$. Beyond this, now we have a simple interpretation
to the mass term: it is just the interaction of the gauge
field with the cosmological constant. Another advantage of our model is the covariance. This enable
us to construct smooth versions of the one proposed in \cite{Ghoroku:2001zu}. This was not possible before since the origin of the delta function was unknown. We show here that our model provides a proof of localizability for a general warp factor.

It has been argued that nonlocalizability happens due to the fact that the gauge field is conformal in four dimension.
Therefore there should be a general way, independent of $A(z)$, for
providing this property. In fact a solution to the Sch\"odinger
equation with potential (\ref{potentialA}) is easily found with the ansatz $\psi(z)=e^{bA(z)}$,
where $b$ is as constant which is fixed to $1/2$ by the equation
of motion and our proof is complete. Note that we have not
used any form of $A(z)$ and our solution is valid for smooth versions
of the RS model. Therefore. this is very powerful to analyze possible solutions
of the localization of gauge fields. With this approach we see, for example, that
a way to solve the problem is by changing consistently both the coefficients
of the two terms in the potential. If now we remember that
\begin{equation}
\label{R}
R=-4(2A''+3A'^{2})e^{-2A}
\end{equation}  
we see that we get for the effective potential
\begin{equation}
\label{effectivepotential}
U=(\frac{1}{4}+12\gamma_{1})A'^{2}+(\frac{1}{2}+8\gamma_{1})A''.
\end{equation}
Pluging this into the Schr\"odinger equation with $m=0$ we get as
before $\gamma_{1}=1/16$ with solution $\psi=e^{A}$. Then our integrand becomes $e^{2A(z)}$ and for any smooth solution such that the RS warp factor is recovered for large $z$ we get a localized zero mode. The correctness of our proof stands in the fact that for any smooth
version of RS model, the above potential is obtained for the transverse
part of the gauge field. Let us prove this now. First we need the equations of motions
\begin{eqnarray}\label{eqmotion}
\partial_{M}(\sqrt{-g}g^{MO}g^{NP}Y_{OP})&=&-\gamma_{1}\sqrt{-g}Rg^{NP}A_{P};
\\\label{transverse}
\quad\partial_{N}(\sqrt{-g}Rg^{NO}A_{O})&=&0,
\end{eqnarray}
where Eq. (\ref{transverse}) is obtained from the
antisymmetry of Eq. (\ref{eqmotion}). The only way to relate this to a massive
gauge field in four dimensions is if we have the condition $\partial_{\mu}X^{\mu}=0$
satisfied. This is surely not true since Eq. (\ref{transverse}) 
give us
\begin{equation}
\label{transverserelation}
e^{3A}R\partial_{\mu}X^{\mu}=-\partial(e^{3A}RX_{5}).
\end{equation}
However, we can split the field in two parts $X^{\mu}=X_{L}^{\mu}+X_{T}^{\mu}$,
where $L$ stands for longitudinal and $T$ stands for transversal
with 
\[
X_{T}^{\mu}=(\delta_{\nu}^{\mu}-\frac{\partial^{\mu}\partial_{\nu}}{\Box})X^{\nu}\;\;\;;\;\;\; X_{L}^{\mu}=\frac{\partial^{\mu}\partial_{\nu}}{\Box}X^{\nu}.
\]

To continue, we must show that the equations above can generate effective
equations for a transverse massive gauge field in four dimensions and
a massive scalar field (zero form) defined by $X_{5}=\Phi$. The first
thing we observe is that Eq. (\ref{transverserelation})
give us a relation between the scalar field and the longitudinal
part of $X^{\mu}$. We will also need the following identities
\begin{eqnarray}
\label{identities}
\partial_{\mu}Y^{\mu\nu}&=&\Box X_{T}^{\nu};\\
Y^{5\mu}&=&\partial X_{T}^{\mu}+\partial X_{L}^{\mu}-\partial^{\mu}\Phi\equiv\partial X_{T}^{\mu}+Y_{L}^{5\mu};\nonumber \\
 Y_{L}^{\mu5}&=&\frac{\partial^{\mu}}{\Box}\partial_{\nu}Y^{\nu5}.\nonumber
\end{eqnarray}
With this, Eq. (\ref{eqmotion}) can be divided in two. We get , for $N=5$ 
\begin{equation}
\label{M=5}
\partial_{\mu}Y^{\mu5}+\gamma_{1} e^{2A}R\Phi=0
\end{equation}
and for $N=\nu$ 
\begin{equation}
\label{M=mu}
e^{A}\Box X_{T}^{\nu}+\partial\left(e^{A}\partial X_{T}^{\nu}\right)+\gamma_{1} e^{3A}RX_{T}^{\nu}+\partial(e^{A}Y_{L}^{5\mu})+\gamma_{1} e^{3A}RX_{L}^{\nu}=0.
\end{equation}
Using (\ref{transverserelation}), (\ref{identities}) and (\ref{M=5}) we get yet
\[
\partial(e^{A}Y_{L}^{\mu5})=-\gamma_{1}\frac{\partial^{\mu}}{\Box}\partial(e^{3A}R\Phi)=-\gamma_{1} e^{3A}RX_{L}^{\nu}
\]
and finally we obtain from Eq. (\ref{M=mu}) the equation for the transverse part of the gauge
field

\[
e^{A}\Box X_{T}^{\nu}+\partial\left(e^{A}\partial X_{T}^{\nu}\right)+\gamma_{1} e^{3A}RX_{T}^{\nu}=0.
\]

Finally, by using  Eq. (\ref{R}) and performing the transformation $\tilde{\psi}=e^{-\frac{A}{2}}\psi$ we get the desired Schr\"odinger equation with potential (\ref{effectivepotential}). For the scalar field we must be careful since we have
\[
\Box\Phi-\partial\partial_{\mu}A^{\mu}-\gamma_{1}Re^{2A}\Phi=0
\]
and performing the separation of variables $\Phi=\Psi(z)\phi(x)$ and the transverse condition we get the equation for the massive mode of the scalar field
\[
\partial(R^{-1}e^{-3A}\partial(Re^{3A}\Psi))-\gamma_{1}Re^{2A}g\Psi=m^{2}\Psi
\]
and defining $g=e^{3A}R$ we get
\begin{equation}\label{equationg}
\partial(g^{-1}\partial(g\Psi))-\gamma_{1}e^{-A}g\Psi=m^{2}\Psi.
\end{equation}
To put the above equation in the Sch\"oedinger form, consider the general equation
\[
\partial(f\partial(g\Psi))+h\Psi=-m^{2}\Psi.
\]
or
\[
-\Psi''-(\ln(fg^{2}))'\Psi'-((fg')'+h)\Psi=m^{2}(fg)^{-1}\Psi^{2},
\]
and comparing with a previous identity found in \cite{Landim:2011ki}
we get $P=-\ln fg^{2}$, $V=-((fg')'+h)$ and $Q=(fg)^{-1}$. With
this we obtain
\[
\frac{dz}{dy}=\Theta(y),\quad\psi(y)=\Omega(y)\overline{\psi}(z),
\]
with
\[
\Theta(y)=(fg)^{-1/2},\quad\Omega(y)=f^{-\frac{1}{4}}g^{-\frac{3}{4}},
\]
and the potential
\[
{U}(z)=V(y)/\Theta^{2}+\left(P'(y)\Omega'(y)-\Omega''(y)\right)/\Omega\Theta^{2}.
\]
With this definition we see that
\[
\Omega'=(\ln f^{-\frac{1}{4}}g^{-\frac{3}{4}})'\Omega;\Omega''=f^{\frac{1}{2}}g^{\frac{3}{2}}(f^{-\frac{1}{4}}g^{-\frac{3}{4}})''\Omega
\]
and we get the final form of the potential as
\begin{eqnarray}
&&U(z)=-((fg')'+h)fg+\\ \nonumber
&&[(\ln f^{-1}g^{-2})'(\ln f^{-\frac{1}{4}}g^{-\frac{3}{4}})'-f^{\frac{1}{4}}g^{\frac{3}{4}}(f^{-\frac{1}{4}}g^{-\frac{3}{4}})'']fg.
\end{eqnarray}
or
\begin{eqnarray}
&&U(z)=-((fg')'+h)fg+\\ \nonumber
&&[(\frac{1}{4}\frac{f''}{f}+\frac{3}{4}\frac{g''}{g}-\frac{1}{16}(\ln f)'^{2}-\frac{1}{16}(\ln g)'^{2}+\frac{5}{4}(\ln f)'(\ln g)']fg.
\end{eqnarray}

In our case we have $f=g^{-1}$ and we get $\Psi=g^{-1/2}\psi$ and
a Schr\"odinger equation with potential
\begin{equation}\label{solutiong}
U=\frac{3}{4}\frac{g'^{2}}{g^{2}}-\frac{1}{2}\frac{g''}{g}+\gamma_{1}e^{-A}g
\end{equation}
or
\[
U=\frac{1}{4}(3A'+(\ln R)')^{2}-\frac{1}{2}(3A''+(\ln R)'')+\gamma_{1}Re^{2A}.
\]

It can  be seing clearly how the equations of motion for the scalar field differ from that of a standard scalar field. Despite the complicated form of the
potential we must note some interesting properties. First, we can see that if we have an asymptotic RS metric we must have $\lim_{z\to\infty}R=-20k^2$ and therefore we get an asymptotic potential
\[
U=\frac{9}{4}A'^{2}-\frac{3}{2}A''+\gamma_{1}(8A''+12A'^{2})
\]
as expected. With this potential we see that the zero mode of the scalar field solution is localized for $\gamma_{1}=9/16$. Showing that we cannot have both fields localized.  At the same time, we have the freedom to choose one of them to localize. But due to the experience in $D=4$, we know that the vector field is the one to be localized. However, the model proposed here does not give any information leading to a natural of the field to be localized.

In this way we have shown that it is possible to localize a zero mode of gauge vector field by regarding the following ingredients: consider a geometrical Yukawa coupling between the gauge fields and the Ricci scalar through a mass term; restoration of gauge symmetry in $D=4$ after breaking due to the mass term in $D=5$; choose, by experience in $D=4$, the gauge vector field to localize.

An interesting issue is the generalization to antisymmetric tensor fields or $p-$forms as in \cite{Alencar:2010vk,Alencar:2010hs,Landim:2010pq}.  In a separate paper we consider this and show that the generalization to $p-$forms in more dimensions is possible. The main result is that the gauge field can emerge from the localization of the two form in five dimensions \cite{Alencar:2014fga}. Another aspect to consider is the generalization to a non-abelian case as in \cite{Batell:2006dp} or to other couplings with the field strength \cite{Germani:2011cv}. The fermionic case with geometrical couplings could be a very interesting case to work and it opens the possibility of finding analytical solutions to this case generalizing the bosonic cases \cite{Alencar:2012en,Landim:2013dja}. The question about physical criterion to choose which field to be localized is very interesting. The model proposed only gives one final free parameter to discuss and we don't see how this question could be addressed now. All of this discussions are left for future work.

\section*{Acknowledgment}
We acknowledge the financial support provided by Funda\c c\~ao Cearense de Apoio ao Desenvolvimento Cient\'\i fico e Tecnol\'ogico (FUNCAP), the Conselho Nacional de 
Desenvolvimento Cient\'\i fico e Tecnol\'ogico (CNPq) and FUNCAP/CNPq/PRONEX.

\providecommand{\href}[2]{#2}\begingroup\raggedright\endgroup
 
\end{document}